\documentclass[a4paper,12pt]{article}
\usepackage{epsfig}
\usepackage{amsmath}
\usepackage{amssymb}
\usepackage{amsfonts}
\usepackage{bbm}
\usepackage{fleqn}
\usepackage[footnotesize]{caption}
\usepackage{graphicx}
\usepackage{mathrsfs}
\usepackage[center,footnotesize,hang]{subfigure}

\def\be{\begin{equation}}
\def\ee{\end{equation}}
\def\beq{\begin{equation}}
\def\eeq{\end{equation}}
\def\bea{\begin{eqnarray}}
\def\eea{\end{eqnarray}}
\def\f{\frac}

\def\<{\left\langle}
\def\>{\right\rangle}

\allowdisplaybreaks[2]
\begin{document}
\bibliographystyle{OurBibTeX}
\begin{titlepage}
 \vspace*{-15mm}
\begin{flushright}
\end{flushright}
\vspace*{5mm}
\begin{center}
{ \sffamily \LARGE Tri-bimaximal-Cabibbo Mixing}
\\[8mm]
S.~F.~King\footnote{E-mail:
\texttt{king@soton.ac.uk}}
\\[3mm]
{\small\it
School of Physics and Astronomy,
University of Southampton,\\
Southampton, SO17 1BJ, U.K.
}\\[1mm]
\end{center}
\vspace*{0.75cm}
\begin{abstract}
\noindent
Recent measurements of the lepton mixing angle $\theta_{13}$
by the Daya Bay and RENO reactor experiments are consistent with the relationship 
$\theta_{13}\approx \theta_C/\sqrt{2}$ where $ \theta_C$ is the Cabibbo angle.
We propose Tri-bimaximal-Cabibbo (TBC) mixing, in which 
$\sin \theta_{13}= \sin \theta_C/\sqrt{2}$, $\sin \theta_{23}= 1/\sqrt{2}$ and $\sin \theta_{12}= 1/\sqrt{3}$.
We show that TBC mixing may arise approximately from Tri-bimaximal, Bi-maximal or Golden Ratio neutrino mixing, 
together with Cabibbo-like charged lepton corrections arising from a Pati-Salam gauge group, leading to 
predictions for the CP-violating phase of $\delta \approx \pm 90^o, \pm 180^o, \pm 75^o$, respectively. 
Alternatively, we show that TBC neutrino mixing may realised accurately using
the type I see-saw mechanism with partially constrained
sequential right-handed neutrino dominance, assuming a family symmetry 
which is broken by a flavon common to quarks and neutrinos.
\end{abstract}
\end{titlepage}
\newpage
\setcounter{footnote}{0}

\section{Introduction}
It is one of the goals of theories of particle physics beyond the
Standard Model to predict quark and lepton masses and mixings,
or at least to relate them.
While the quark mixing angles are known to all be rather small, by contrast two of the lepton mixing angles,
the atmospheric angle $\theta_{23}$ and the solar angle $\theta_{13}$,
are identified as being rather large \cite{Nakamura:2010zzi}. This is usually interpreted as demonstrating that quark 
mixing is very different from lepton mixing. However 
the smallest remaining angle, the reactor angle 
$\theta_{13}$, has recently been measured and its value shown to be not that small.
We shall discuss the implications of the observation that the smallest lepton mixing angle, $\theta_{13}$,
may be related to the largest quark mixing angle, the Cabibbo angle, $\theta_C$,
providing a possible link between lepton and quark mixing.

Early indications for $ \theta_{13}$ from global fits 
were given in \cite{Fogli:2008jx}.
Direct evidence for $\theta_{13}$
was first provided by T2K, MINOS and Double CHOOZ
\cite{Abe:2011sj,Adamson:2011qu,Abe:2011fz}.
Global fits including these results were subsequently given in
 \cite{Fogli:2011qn,Schwetz:2011zk}.
Recently, Daya Bay \cite{An:2012eh} have measured,
\be
\sin^2 2  \theta_{13}=0.092\pm 0.016 \, \mathrm{(stat.)} \pm 0.005 \, \mathrm{(syst.)},
\ee
while, shortly afterwards,
RENO \cite{Ahn:2012nd} have measured,
\be
\sin^2 2  \theta_{13}=0.113\pm 0.013 \, \mathrm{(stat.)} \pm 0.019 \, \mathrm{(syst.)}.
\ee
The global fits for the solar and atmospheric in 
\cite{Fogli:2011qn,Schwetz:2011zk}, together with the above Daya Bay and RENO
results for the reactor angle, lead to the approximate one sigma ranges for
the lepton mixing angles,
\beq
\theta_{13}= 9^o\pm 1^o, \ \
\theta_{12}= 34^o\pm 1^o, \ \
\theta_{23}= {45^o}\pm 5^o.
\label{fits}
\eeq

The above results from the Daya Bay and RENO reactor experiments are consistent with a remarkable relationship
between the smallest lepton mixing angle $\theta_{13}$
and the largest quark mixing angle, $\theta_C$, namely $\theta_{13}\approx \theta_C/\sqrt{2}$,
where the Cabibbo angle $ \theta_C\approx 13^o$ implies $\theta_{13}\approx 9.2^o$. 
In section \ref{sect:TBC} we combine this relation
with maximal atmospheric mixing and trimaximal solar mixing to give
Tri-bimaximal-Cabibbo (TBC) mixing in which 
$\sin \theta_{13}= \sin \theta_C/\sqrt{2}$, $\sin \theta_{23}= 1/\sqrt{2}$ and $\sin \theta_{12}= 1/\sqrt{3}$.
In section \ref{sect:strategy} we show how
approximate TBC mixing may emerge from Tri-bimaximal, Bi-maximal or Golden Ratio neutrino mixing, by invoking Cabibbo-like charged lepton corrections, leading to approximate predictions for the CP-violating phase of $\delta \approx \pm 90^o, \pm 180^o, \pm 75^o$, respectively.
The required Cabibbo-like charged lepton mixing
may be present in Pati-Salam models with a particular Clebsch structure.
Alternatively, in section \ref{sect:PCSD}, we show how accurate TBC neutrino mixing may arise from
the type I see-saw mechanism with partially constrained
sequential right-handed neutrino dominance. 
This may be realised in models with a family symmetry
where a misaligned flavon is common to both the neutrino and quark sectors.
The summary and conclusions are given in section \ref{sect:conclusions}.

\section{Tri-bimaximal-Cabibbo Mixing \label{sect:TBC}}

The recent data  is consistent with the remarkable relationship,
\beq
s_{13}= \frac{ \sin \theta_C }{\sqrt{2}} = \frac{\lambda}{\sqrt{2}}, 
\label{rem1}
\eeq
where $\lambda  = 0.2253\pm 0.0007$ \cite{Nakamura:2010zzi} 
is the Wolfenstein parameter. 
This relationship is an example of ``Cabibbo Haze'' \cite{Datta:2005ci}, 
the general hypothesis that the Cabibbo angle is an expansion parameter for lepton as
well as quark mixing. It was proposed earlier in the context of ``Quark-Lepton Complementarity'' (QLC)
in which $\theta_{12}+\theta_C = 45^o$ \cite{Minakata:2004xt}. For related approaches see \cite{related}.
Our approach in section \ref{sect:strategy} relies on maximal atmospheric mixing but the solar angle is 
determined by ``Sum Rules'' \cite{sumrule}, which differ from the QLC relation. 
These examples illustrate that the value of the solar angle is independent of the relation in Eq.\ref{rem1}. 
On the other hand, phenomenology 
is consistent with a trimaximal solar angle as in Eq.\ref{fits}, and furthermore 
the approach in section \ref{sect:PCSD} suggests a trimaximal solar angle.
It is therefore natural to combine Eq.\ref{rem1} with TB mixing, as discussed below.

In terms of the combination measured by the reactor neutrino experiments,
Eq.\ref{rem1} implies,
\beq
\sin^22\theta_{13}\approx 2\lambda^2(1-\frac{\lambda^2}{2})\approx 0.099,
\eeq
in excellent agreement with the recent Daya Bay and RENO results above.
Furthermore the above ansatz implies a reactor angle of 
\beq
\theta_{13}\approx \frac{\theta_C}{\sqrt{2}}\approx 9.2^o,
\label{rem2}
\eeq
where $\theta_C\approx 13^o$ is the Cabibbo angle.

Apart from the reactor angle, the measured and fitted atmospheric and solar angles are in good agreement with the ansatz of Tri-bimaximal (TB) mixing \cite{HPS}. We are therefore led to combine the relation
in Eq.\ref{rem1}
with TB mixing to yield tri-bimaximal-Cabibbo (TBC) mixing:
\be
s_{13} = \frac{\lambda}{\sqrt{2}}, \ \ s_{12} = \frac{1}{\sqrt{3}},
\ \ s_{23} = \frac{1}{\sqrt{2}}.
\label{TBC0}
\ee
In terms of the TB deviations parameters defined in \cite{King:2007pr}, this corresponds to 
$r=\lambda$ with $s=a=0$. Using the second order expansion in \cite{King:2007pr}, Eq.\ref{TBC0}
then leads to the following approximate form of the mixing matrix, 
\begin{eqnarray}
U_{TBC} \approx
\left( \begin{array}{ccc}
\sqrt{\frac{2}{3}}(1-\frac{1}{4}\lambda^2)  & \frac{1}{\sqrt{3}}(1-\frac{1}{4}\lambda^2) 
& \frac{1}{\sqrt{2}}\lambda e^{-i\delta } \\
-\frac{1}{\sqrt{6}}(1+\lambda e^{i\delta })  & \frac{1}{\sqrt{3}}(1- \frac{1}{2}\lambda e^{i\delta })
& \frac{1}{\sqrt{2}}(1-\frac{1}{4}\lambda^2) \\
\frac{1}{\sqrt{6}}(1- \lambda e^{i\delta })  & -\frac{1}{\sqrt{3}}(1+ \frac{1}{2}\lambda e^{i\delta })
 & \frac{1}{\sqrt{2}}(1-\frac{1}{4}\lambda^2)
\end{array}
\right)P + \mathcal{O}(\lambda^3),
\label{TBC}
\end{eqnarray}
corresponding to the mixing angles,
\beq
\theta_{13}\approx 9.2^o, \ \ 
\theta_{12}= 35.26^o, \ \
\theta_{23}= 45^o. \ \
\label{TBangles}
\eeq

\section{TBC mixing from Charged Lepton Corrections  \label{sect:strategy}}

In a typical convention (see e.g. \cite{King:2002nf}) the PMNS matrix may be constructed as,
\be
U =U^e{U^{\nu}}^{\dagger}
\ee
where in models, $U^e$ is related to the left-handed rotations involved in diagonalising
the charged lepton mass matrix $M^e$, and $U^{\nu}$ is related to the matrix that diagonalises the
left-handed Majorana neutrino mass matrix $m^{\nu}$. A particular model typically has a preference for a particular
basis in which $M^e$ and $m^{\nu}$ take a particular form, leading to $U^e$ and $U^{\nu}$ having also particular forms which may be separately parameterised by $e$ and $\nu$ mixing angles in analogy with the PDG parameterisation. Here we show how Eq.\ref{rem2} can simply arise from a zero neutrino mixing angle
$\theta_{13}^\nu \approx  0$ with Cabibbo-like charged lepton corrections $\theta_{12}^e \approx \theta_C$.

In many grand unified theories (GUTs) the charged lepton mixing angles are dominated by 
$\theta_{12}^e  \gg \theta_{13}^e,  \theta_{23}^e$.
Furthermore, assuming $\theta_{12}^e \gg \theta_{13}^{\nu}$, it has been widely observed that charged lepton corrections then imply \cite{King:2002nf,ChargedLeptonCorrections},
\be
\theta_{13} \approx \frac{\theta_{12}^e}{\sqrt{2}}.
\label{1312}
\ee
Note that the factor of $1/\sqrt{2}$ arises from maximal atmospheric neutrino mixing.
In order to achieve Eq.\ref{rem2} we need only assume that 
the dominant charged lepton angle $\theta_{12}^e$ is equal to the Cabibbo angle,
\be
 \theta_{12}^e \approx \theta_C .
 \label{12}
\ee
Then Eqs.\ref{1312} and \ref{12} imply,
\beq
\theta_{13}\approx \frac{\theta_C}{\sqrt{2}}\approx 9.2^o,
\label{rem3}
\eeq
in agreement with Eq.\ref{rem2}.

\subsection{Simple neutrino mixing patterns}
Below we give three classic examples of simple patterns of mixing in the neutrino sector which all
have $\theta_{13}^{\nu}=0$ and $\theta_{23}^{\nu}=45^o$, namely tri-bimaximal (TB) neutrino mixing
\cite{HPS}, bi-maximal (BM) neutrino mixing
(see e.g. \cite{Davidson:1998bi, Altarelli:2009gn} and references therein), 
and the Golden Ratio (GR) neutrino mixing \cite{Datta:2003qg, Everett:2008et}.
They all lead to solar mixing angle Sum Rules \cite{sumrule} involving the 
physical CP violating oscillation phase $\delta$  \cite{Nakamura:2010zzi}, 
\be
\theta_{12}\approx \theta_{12}^{\nu}+\frac{\theta_C}{\sqrt{2}}\cos \delta ,
\label{sumrule}
\ee
where $\theta_{12}^{\nu}=35.26^o, 45^o, 31.7^o$ for the case of TB, BM, GR 
\footnote{Note that there is an alternative version of GR mixing where 
$\cos \theta_{13}^{\nu} =\phi/2$ and $\theta_{12}^{\nu}=36^o$ \cite{Rodejohann:2008ir}.
This leads to the Sum Rule $\theta_{12}\approx 36^o+ 9.2^o\cos \delta$,
numerically similar to the case of TB neutrino mixing.}
neutrino mixing respectively, where we have used the prediction in Eq.\ref{rem3}.
Note that the Sum Rule is subject to (typically small) corrections due to renormalisation group running and 
canonical normalisation effects \cite{Antusch:2008yc}.
Given the prediction in Eq.\ref{rem3} for $\theta_{13}$, the sum rule 
in Eq.\ref{sumrule} then yields
a favoured range of $\cos \delta $ in each case which may be tested 
in proposed neutrino experiments \cite{Antusch:2007rk}.
We discuss this in more detail below for each of the three
cases:
\vspace{0.2in}

(i) \underline{TB mixing in the neutrino sector}\\
\be
s_{13}^{\nu} = 0, \ \ s_{12}^{\nu} = \frac{1}{\sqrt{3}},
\ \ s_{23}^{\nu} = \frac{1}{\sqrt{2}},
\label{TBnu}
\ee
together with with charged lepton corrections 
\be
 \theta_{12}^e \approx \theta_C \approx \lambda \gg \theta_{13}^e,  \theta_{23}^e
\ee
yields to first order in $ \theta_{12}^e$ \cite{sumrule}, 
\be
s_{13} \approx \frac{\lambda}{\sqrt{2}}, \ \ s_{12} \approx \frac{1}{\sqrt{3}} + \frac{\lambda}{\sqrt{3}}\cos \delta,
\ \ s_{23} \approx \frac{1}{\sqrt{2}}.
\label{TBCapprox}
\ee
This is not quite of the TBC form in Eq.\ref{TBC0} due to the large deviation in the solar angle,
leading to the approximate linear relation between the solar angle and $\cos \delta$,
\be
\theta_{12}\approx 35.26^o+ 9.2^o\cos \delta .
\ee
Thus TB neutrino mixing implies that $\cos \delta \approx 0$ 
or $\delta \approx \pm 90^o$ in order that the solar angle
does not deviate to much from its TB value.

\vspace{0.2in}

(ii)  \underline{Bi-maximal neutrino mixing}
\be
s_{13}^{\nu} = 0, \ \ s_{12}^{\nu} = \frac{1}{\sqrt{2}},
\ \ s_{23}^{\nu} = \frac{1}{\sqrt{2}},
\label{BMnu}
\ee
together with with charged lepton corrections 
\be
 \theta_{12}^e \approx \theta_C \approx \lambda \gg \theta_{13}^e,  \theta_{23}^e
\ee
yields to first order in $ \theta_{12}^e$  \cite{sumrule}, 
\be
s_{13} \approx \frac{\lambda}{\sqrt{2}}, \ \ s_{12} \approx \frac{1}{\sqrt{2}} + \frac{\lambda}{2}\cos \delta,
\ \ s_{23} \approx \frac{1}{\sqrt{2}}.
\label{TBCapprox}
\ee
Again this is not quite of the TBC form in Eq.\ref{TBC0} due to the large deviation in the solar angle,
leading to the approximate linear relation between the solar angle and $\cos \delta$,
\be
\theta_{12}\approx 45^o+9.2^o\cos \delta .
\ee
Thus BM neutrino mixing implies that $\cos \delta \approx -1$ or $\delta \approx \pm 180^o$
 in order to achieve a solar angle
$\theta_{12}\approx 36^o$. Note that this is a very distinct prediction from the above case of TB neutrino mixing
where we predict $\delta \approx \pm 90^o$.
\vspace{0.2in}

(ii)  \underline{Golden Ratio neutrino mixing}
\be
s_{13}^{\nu} = 0, \ \  s_{12}^{\nu} = \frac{1}{\sqrt{1+\phi^2}},
\ \ s_{23}^{\nu} = \frac{1}{\sqrt{2}},
\label{BMnu}
\ee
where the Golden Ratio is $\phi = (1+\sqrt{5})/2$,
together with charged lepton corrections 
\be
 \theta_{12}^e \approx \theta_C \approx \lambda \gg \theta_{13}^e,  \theta_{23}^e
\ee
yields to first order in $ \theta_{12}^e$ the relation, 
\be
s_{13} \approx \frac{\lambda}{\sqrt{2}}, \ \ 
s_{12} \approx  \frac{1}{\sqrt{1+\phi^2}} + \frac{\lambda}{\sqrt{2}}\frac{\phi}{\sqrt{1+\phi^2}}\cos \delta,
\ \ s_{23} \approx \frac{1}{\sqrt{2}}.
\label{TBCapprox}
\ee
As before, this is not quite of the TBC form due to the deviation in the solar angle,
leading to the approximate linear relation between the solar angle and $\cos \delta$,
\be
\theta_{12}\approx 31.7^o+9.2^o\cos \delta .
\ee
Thus GR neutrino mixing implies that $\cos \delta \approx 0.25$ 
or $\delta \approx \pm 75^o$ in order to achieve a solar angle
$\theta_{12}\approx 34^o$.
This is closer to the case of TB neutrino mixing
where we predict $\delta \approx \pm 90^o$, but is very different from BM neutrino mixing
where we predict $\delta \approx \pm 180^o$.

\subsection{Models with $ \theta_{12}^e \approx \theta_C$ }
A crucial assumption of the above approach is that the dominant charged lepton
mixing angle is equal to the Cabibbo angle, namely that $ \theta_{12}^e \approx \theta_C$,
as stated in Eq.\ref{12}.
This provides a connection between the charged lepton sector and the 
quark sector which may hint at some underlying quark-lepton unification.
However traditional quark-lepton unification models involve different relations,
for example the Georgi-Jarlskog (GJ) \cite{GJ} prediction would be $ \theta_{12}^e \approx \theta_C/3$.
Other more recent studies which consider large values of $ \theta_{12}^e$ in GUT models
\cite{Antusch:2011qg,Marzocca:2011dh} do not explain $ \theta_{12}^e \approx \theta_C$.
It therefore requires some discussion about how this might be achieved in models. 

Let us focus on the upper $2\times 2$
block of the mass matrices, 
assuming an approximately diagonal up type quark Yukawa matrix.
In order to achieve $ \theta_{12}^e \approx \theta_C$ we propose a structure
 (in a LR convention for mass matrices):
\be
Y^d_{2\times 2}\propto
 \begin{pmatrix} 
  \ast & c_d\lambda \\
  \ast  & c_d
 \end{pmatrix} , \ \ Y^e_{2\times 2}\propto
 \begin{pmatrix} 
  \ast & c_e\lambda \\
   \ast  & c_e\end{pmatrix} 
\ee
where the factors of $c_{e,d}$ represents the effect of Clebch coefficients in some unified model
where the assumption of equal Clebsch factor in the (1,2) and (2,2) elements leads to the 
relation $ \theta_{12}^e \approx \theta_C$.
For example the choice $c_e/c_d=3$ gives the mass relation $m_{\mu}= 3m_s$ at the GUT scale.
Note that there are many ways to obtain the correct electron and down quark masses, which depend on the unspecified elements denoted by ``$\ast $'' above (assumed to be smaller than the (2,2) element). 
For example, operators exist which contribute to 
either the charged lepton or the down quark mass matrix, but not both at the same time, 
in Pati-Salam models \cite{Allanach:1996hz,Antusch:2005ca}.

It is not the purpose of this paper to provide a detailed model, but the general strategy is clear.
One may start from some family symmetry $G_F$ which is capable of yielding a simple pattern
of neutrino mixing such as TB, BM or GR, for example $G_F=A_4, S_4, A_5$
as discussed in many papers (for a review see e.g.\cite{Altarelli:2010gt}). Then one must extend such 
a model to include both the quarks and leptons by assuming a Pati-Salam gauge group, for example, with 
mass matrices of the above form, leading to Cabibbo-like charged lepton corrections.
We emphasise that the key feature of this approach is a Cabibbo-like charged lepton correction 
$\theta_{12}^e \approx \theta_C$, starting from a zero neutrino mixing angle $\theta_{13}^\nu \approx  0$.
Other approaches to obtain a large reactor angle are discussed in \cite{bulk}.

\section{\label{sect:PCSD} TBC mixing in the Neutrino Sector  }

In the previous section we showed how Eq.\ref{rem1} could arise in cases with 
$\theta_{13}^\nu \approx  0$ and $\theta_{12}^e \approx \theta_C$.
In this section we show how it can arise in models with zero charged lepton corrections,
namely $\theta_{13}^\nu \approx  \theta_C/\sqrt{2}$ and $\theta_{12}^e \approx 0$.
In order to achieve this we need to explain two things: (i) the appearance of the Cabibbo angle in the
neutrino sector, (ii) the factor of $\sqrt{2}$. In this section we show that these features may arise starting from
the type I see-saw mechanism \cite{Minkowski:1977sc}
with sequential right-handed neutrino dominance (SD) \cite{King:1998jw}.

In the first subsection below we start by reviewing SD and show how it can give the factor of 
$\sqrt{2}$ in Eq.\ref{rem1}. Although SD is well known, it is instructive to go through these arguments
to see how the factor of $\sqrt{2}$ arises in $\theta_{13}$ 
from maximal atmospheric neutrino mixing, and to set the
scene for the vacuum alignments which follow.
In the subsequent subsection we show how 
these assumptions may be justified dynamically starting from a family symmetry, based on 
symmetry breaking flavons with particular vacuum alignments.
In the final subsection we show how
one of the flavons involving $\lambda$ must appear in both the neutrino and quark sectors
in order to account for both the reactor angle and Cabibbo mixing in the quark sector.

\subsection{Sequential right-handed neutrino dominance}
First consider the 
case of single right-handed neutrino dominance where only one right-handed neutrino $N_3^c$
of heavy Majorana mass $M_3$
is present in the see-saw mechanism, namely the one responsible for 
the atmospheric neutrino mass $m_3$ \cite{King:1998jw}. If the single right-handed neutrino
couples to the three lepton doublets $L_i$ in the diagonal charged lepton mass basis as,
\be
H (dL_e + eL_{\mu}+ f L_{\tau}) N_3^c,
\ee
where $d,e,f$ are Yukawa couplings (assumed real for simplicity
\footnote{The full results including phases are discussed in \cite{King:1998jw}.}), and $H$ is the Higgs doublet,
where it is assumed that $d\ll e,f$. so that the see-saw mechanism yields
the atmospheric neutrino mass, 
\be
m_3\approx (e^2+f^2)\frac{v^2}{M_3},
\label{m3}
\ee
where 
$v=\langle H\rangle$.
Then the reactor and atmospheric angles are approximately given by 
simple ratios of Yukawa couplings \cite{King:1998jw},
\be
\theta_{13}\approx \frac{d}{\sqrt{e^2+f^2}}, \ \  \tan \theta_{23}\approx \frac{e}{f}.
\label{angles1}
\ee
If the Yukawa couplings would satisfy the condition,
\be
 \begin{pmatrix} 
  d\\e\\f \end{pmatrix} = a_3
  \begin{pmatrix} 
  \lambda \\1 \\1 \end{pmatrix}, 
  \label{cond1}
\ee
then Eqs.\ref{angles1} and \ref{cond1} imply,
\be
\theta_{13}\approx \frac{\lambda }{\sqrt{2}}, \ \  \tan \theta_{23}\approx 1
\label{predictions1}
\ee
which predicts the desired relation in Eq.\ref{rem1}, with the 
factor of $\sqrt{2}$ arising from maximal atmospheric mixing.
However we need to show that this relation is not spoiled when
other right-handed neutrinos are included.

According to sequential dominance (SD) \cite{King:1998jw}
the solar neutrino mass and mixing are accounted for by introducing 
a second right-handed neutrino $N_2^c$ with mass $M_2$ which
couples to the three lepton doublets $L_i$ in the diagonal charged lepton mass basis as,
\be
H (aL_e + bL_{\mu}+ c L_{\tau}) N_2^c,
\ee
where $a,b,c$ are Yukawa couplings (assumed real for simplicity). Then the second right-handed neutrino
is mainly responsible for the solar neutrino mass, providing 
\be
(a,b,c)^2/M_2\ll (e,f)^2/M_3,
\ee
which is the basic SD condition. Assuming this, then the see-saw mechanism leads to the solar
neutrino mass,
\be
m_2\approx \left(a^2+ (c_{23}b-s_{23}c)^2\right)\frac{v^2}{M_2},
\label{m2}
\ee
and the solar neutrino mixing is approximately given by 
a simple ratios of Yukawa couplings \cite{King:1998jw},
\be
\tan \theta_{12}\approx \frac{a}{(c_{23}b-s_{23}c)}.
\label{angles2}
\ee
So far the subdominant Yukawa couplings $a,b,c$ are unconstrained.
The presence of these couplings will 
in general affect the reactor angle and destroy the relation in Eq.\ref{rem1},
since there is an additional contribution of the form \cite{King:1998jw},
\be
\Delta \theta_{13}  \approx 
\frac{a(eb+fc)}{(e^2+f^2)^{3/2}}
\frac{M_3}{M_2}
\label{Delta}
\ee
It is clear that, if the lower two components of the two Yukawa column vectors 
are orthogonal (i.e. if $eb+fc=0$) then the reactor angle will be unaffected.
Choosing the subdominant Yukawa couplings to satisfy,
\be
 \begin{pmatrix} 
  a\\b\\c \end{pmatrix} =a_2
  \begin{pmatrix} 
  1 \\1 \\-1 \end{pmatrix}, 
  \label{cond2}
\ee
leaves the reactor angle unchanged from its value in Eq.\ref{predictions1}, 
since $\Delta \theta_{13} =0$, while 
Eqs.\ref{angles2} and \ref{cond2}, together with maximal atmospheric mixing,
implies trimaximal solar mixing,
\be
\tan \theta_{12}\approx \frac{1}{\sqrt{2}}.
\label{predictions2}
\ee
Eqs.\ref{predictions1} and \ref{predictions2} realise
TBC mixing as in Eq.\ref{TBC0}, assuming no charged lepton corrections.

Note that, in the limit $\lambda \rightarrow 0$,
Eqs.\ref{cond1} and \ref{cond2} are just the conditions of constrained sequential dominance (CSD) 
where the precise orthogonality of the columns was responsible for $\theta_{13}=0$
\cite{King:2005bj}, leading to TB mixing.
Eqs.\ref{cond1} and \ref{cond2} with a general parameter $\epsilon$
instead of $\lambda$ is referred to as
partially constrained sequential dominance (PCSD), leading to 
tri-bimaximal-reactor (TBR) mixing \cite{King:2009qt}.
Although the above argument is only valid to leading order in $\lambda$,
and $m_2/m_3$, TBR mixing is in fact realised much more accurately than this as shown 
numerically \cite{arXiv:1003.5498} and analytically \cite{King:2011ab}.

\subsection{Family symmetry and flavons in the neutrino sector}
In order to account for the equality
of Yukawa couplings in Eqs.\ref{cond1} and \ref{cond2} such as $a=b=-c$ and $e=f$,
we need to introduce
a family symmetry $G_F$ which is broken by flavons $\varphi$ with particular vacuum
alignments. 

Let us therefore introduce a discrete family symmetry $G_F$
which is broken by the VEVs of triplet flavon fields which are aligned as follows,
\be
\langle \varphi_{3} \rangle =  
v_{3} \begin{pmatrix} \lambda \\1\\1 \end{pmatrix} ,
\qquad
\langle \varphi_{2} \rangle = 
v_{2} \begin{pmatrix} 
  1\\1\\-1 \end{pmatrix}  .
\label{a4-align-nu}
\ee
The idea is that these flavons are responsible for generating the columns of the 
neutrino Yukawa matrix in Eqs.\ref{cond1} and \ref{cond2}, associated with the right-handed
neutrinos $N_3^c$ and $N_2^c$, respectively.

For example, consider
a very simple type I see-saw model, based on $G_f = A_4$, as discussed in 
\cite{King:2011ab}. In this model we identify the left-handed lepton doublets $L$ and flavons $\varphi_{i}$ with
$A_4$ triplets, while the right-handed neutrinos $ N_i^c$ and Higgs doublets $H_u$ are $A_4$ singlets.
The neutrino part of the effective Lagrangian reads,
\bea
{\cal L}^{\nu } &\sim & \sum_{i=2}^3\left(
LH_u \frac{\varphi_{i}}{M_{\chi_i}} N_i^c
+ 
N^c_iN^c_i \frac{\varphi_{i}\varphi_{i}}{M_{\Upsilon_{i}}} \right)  ,
\label{a4-Ynu0}
\eea
where the mixing term
$N_2^cN_3^c\varphi_{2}\varphi_{3}$ is forbidden by a choice of
appropriate messengers \cite{King:2011ab}. 

Inserting the Higgs and flavon VEVs, whose alignment is discussed in \cite{King:2011ab},
 leads to the Dirac and
 right-handed Majorana neutrino mass matrices below, 
\be
m_D~=~\begin{pmatrix} a_2&a_3\, \lambda  \\a_2 & a_3\\-a_2 & a_3 \end{pmatrix}v \ ,
\qquad
M_R~=~ \begin{pmatrix} M_2 &0\\0&M_3\end{pmatrix}
\ .\label{m}
\ee
The Dirac mass matrix $m_D$ in Eq.\ref{m}, emerging from a family symmetry, leads to the same result as
the previous assumption that the Yukawa couplings satisfy the constraints in Eqs.~\ref{cond1}, \ref{cond2}.
The family symmetry therefore provides a justification for the previous assumption.

Using the type~I seesaw formula we can express the light neutrino mass matrix as
\be
m_\nu^{}
= 
m_D^{} M_R^{-1} m_D^T
\approx  
\frac{m_2}{3} \begin{pmatrix}
1&1&-1 \\
1&1&-1\\
-1 &-1&1
\end{pmatrix}
+
\frac{m_3}{2} \begin{pmatrix}
 \lambda^2&\lambda&\lambda\\
\lambda&1&1\\
\lambda &1&1
\end{pmatrix},
\label{TBCmassmatrix}
\ee
where we have written $m_3/2\approx a_3^2v^2/M_3$ and $m_2/3\approx a_2^2v^2/M_2$ 
from Eqs.\ref{m3} and \ref{m2}.
Eq.\ref{TBCmassmatrix} leads to TBC mixing in Eq.~\ref{TBC0},
with the deviations being 
of order $\lambda^2$ (where $m_2/m_3\sim \lambda$)
multiplied by small coefficients, as discussed in \cite{King:2011ab}.
This means that the  order $\lambda^2$ corrections to the mixing matrix
closely approximate to those shown in Eq.\ref{TBC}.

\subsection{The charged fermion sector} 
In order to account for the appearance of the Wolfenstein parameter $\lambda$ 
in Eq.\ref{cond1} it is necessary that the flavon involving $\lambda$ should
be common to both the quark and neutrino sectors. We also need to justify that the
charged lepton Yukawa matrix is diagonal.

In order to account for the diagonal charged lepton Yukawa structure
we identify the right-handed charged leptons $e^c,  \mu^c, \tau^c$ with
$A_4$ singlets, and distinguish them using $Z_4$ symmetries
as discussed in \cite{King:2011ab}. 
The resulting effective charged lepton Lagrangian then takes the form
\be
{\cal L}^e ~\sim ~ \frac{1}{M_{\Omega_e}} H_d 
\left( L \varphi_\tau  \tau^c + L \varphi_\mu  \mu^c + L \varphi_e  e^c \right)
 \ .\label{a4-yuk-eff}
\ee
Inserting the flavon VEVs
\be
\langle \varphi_\tau \rangle = v_\tau 
\begin{pmatrix} 0\\0\\1 \end{pmatrix} \ , \qquad
\langle \varphi_\mu \rangle =v_\mu 
\begin{pmatrix} 0\\1\\0 \end{pmatrix} \ ,\qquad
\langle \varphi_e \rangle =v_e 
\begin{pmatrix} 1\\0\\0 \end{pmatrix}
 \ ,\label{a4-align-char0}
\ee
whose alignment is discussed in  \cite{King:2011ab}, leads to
\be
{\cal L}^e ~\sim ~ \frac{1}{M_{\Omega_e}} H_d 
\left(v_\tau L_3 \tau^c + v_\mu L_2 \mu^c + v_e L_1 e^c \right)
 \ , \label{hierarchy}
\ee
thus yielding a diagonal charged lepton mass matrix, which justifies ignoring charged lepton
mixing angle corrections. In a more realistic model, the charged lepton mass hierarchy may be
accounted for via the Froggatt-Nielsen mechanism \cite{CERN-TH-2519}.

The quark sector was not discussed in \cite{King:2011ab}.
Here we suppose that the up-type quark mass matrix is diagonal, as for the charged leptons,
which may be achieved by treating the quark doublets $Q$ as $A_4$ triplets 
and the right-handed up-type quark singlets $U^c$ as $A_4$ singlets, 
and using the same flavons as in the charged lepton sector,
\be
{\cal L}^{u} ~\sim ~ \frac{1}{M_{\Omega_u}} H_u 
\left( Q \varphi_\tau  t^c + Q \varphi_\mu  c^c + Q \varphi_e  u^c \right)
 \ .\label{a4-u-yuk-eff}
\ee
Inserting the flavon VEVs
leads to
\be
{\cal L}^{u} ~\sim ~ \frac{1}{M_{\Omega_u}} H_u 
\left(v_\tau Q_3 t^c + v_\mu Q_2 c^c + v_e Q_1 u^c \right)
 \ , \label{hierarchy-u}
\ee
where, as before, the up-type quark mass hierarchy may be
accounted for via the Froggatt-Nielsen mechanism \cite{CERN-TH-2519}.

The quark mixing must arise from the down-type quark mass matrix. 
In order to achieve this, the right-handed down-type quark singlets $D^c$ are assigned as $A_4$ triplets.
The down-type quark Lagrangian takes the form, assuming that the diagonal contraction
$Q.D^c$ is forbidden by suitable messenger arguments,
\footnote{The messengers responsible for these operators
are required to couple to the bilinear pairs  $(Q .\varphi_{b})$, $( \varphi_{b}.D^c)$,
$(Q .\varphi_{3})$, $( \varphi_{3}.D^c)$ which means that the messengers $\Omega_3$ and
$\Omega_b$ are required to be colour triplets and antitriplets, which forbids them from coupling to $(Q.D^c)$. 
The flavon $\varphi_{b}$ and messenger $\Omega_b$ are odd under a $Z_2^b$ symmetry,
such that $\varphi_{b}$ only couples to the messenger $\Omega_b$
which is coloured and does not enter the lepton sector.
The messenger mass is generated by a field $S^b$ which is odd under $Z_2^b$, giving
$\langle S^b \rangle \overline{\Omega_b}\Omega_b$.}

\bea
{\cal L}^{d} ~&\sim &~ 
\frac{1}{M^2_{\Omega_{bb}}} H_d
(Q .\varphi_{b})( \varphi_{b}.D^c) 
+ \frac{1}{M^2_{\Omega_{33}}} H_d 
(Q .\varphi_{3})( \varphi_{3}.D^c)  
\nonumber \\
&+&\frac{1}{M^2_{\Omega_{3b}}} H_d 
(Q .\varphi_{3})( \varphi_{b}.D^c) + 
\frac{1}{M^2_{\Omega_{b3}}} H_d
(Q .\varphi_{b})( \varphi_{3}.D^c) 
 \ ,\label{a4-yuk-down}
\eea
where $\varphi_{b}$ is a triplet flavon with an alignment,
\be
\langle \varphi_{b} \rangle = 
v_{b} \begin{pmatrix} 
  0\\\delta^2\\1 \end{pmatrix} ,
  \ee
 which may be achieved in a similar way to the flavon $\varphi_{\tau}$
 responsible for the tau lepton mass \cite{King:2011ab}, here allowing for a small misalignment
 $\delta^2\sim {\cal O}(m_s/m_b)$.
 Note that the flavon $\varphi_{3}$ from the neutrino sector, which is responsible for 
 the reactor angle, also appears in Eq.\ref{a4-yuk-down} and is responsible for Cabibbo mixing,
 as follows. 
 
 The down-type quark mass matrix arising from Eq.\ref{a4-yuk-down} is,
 \be
M^d
= 
m_{bb} \begin{pmatrix}
0 & 0 & 0 \\
0 & \delta^4 & \delta^2 \\
0 & \delta^2 &1
\end{pmatrix}
+
m_{33}\begin{pmatrix}
 \lambda^2&\lambda&\lambda\\
\lambda&1&1\\
\lambda &1&1
\end{pmatrix}
+
m_{3b} \begin{pmatrix}
 0 & 0 &\lambda\\
 0 & 0 &1\\
 0 & 0 &1
\end{pmatrix}
+
m_{b3} \begin{pmatrix}
 0 & 0 & 0 \\
 0 & 0 & 0 \\
\lambda &1&1
\end{pmatrix},
\label{downmassmatrix}
\ee
where we assume that the mass parameters 
$m_{\alpha \beta}\sim v_d \< \varphi_{\alpha}  \> \< \varphi_{\beta}  \>/M^2_{\Omega_{\alpha \beta}}$ 
satisfy 
$m_{bb}\approx m_b$ 
and $m_{33}\approx m_s/2$ with $m_{3b}, m_{b3} \leq m_{33}$,
leading to $m_d \sim \lambda^2(m_{3b}-m_{b3})$, since the determinant of the mass matrix vanishes in the limit
$m_{3b}=m_{b3}$. 

The fact that we have used the same flavon
$\varphi_{3}$ in the down sector as in the neutrino sector
implies the relation, 
\be
V_{us} =  \lambda + {\cal O}(\lambda \delta^2),
\ee
which verifies that the Wolfenstein parameter $\lambda$ yields 
Cabibbo mixing in the quark sector. In other words, the same parameter
that arises in the first component of the vacuum alignment of $\varphi_{3}$ is responsible
for both neutrino reactor mixing and Cabibbo mixing, and we identify this parameter with the
Wolfenstein parameter $\lambda$, leading to the relation in Eq.\ref{rem1}.
Assuming $m_{bb}\delta^2 \sim m_{33}$, the mass matrix $M^d$ also implies
$V_{ub}/V_{cb}\sim \lambda/2$, 
in reasonable agreement with experiment \cite{Nakamura:2010zzi}.

\section{Summary and Conclusion \label{sect:conclusions}}
To summarise, recent data from the Daya Bay and RENO reactor experiments is consistent with a remarkable relationship between the smallest lepton mixing angle, $\theta_{13}$, and the largest quark mixing angle,
$\theta_C$, namely $\theta_{13}\approx \theta_C/\sqrt{2}$.
We have proposed a new mixing ansatz called Tri-Bimaximal-Cabibbo (TBC) mixing 
which combines this relation with TB atmospheric and solar mixing. 
We then discussed two ways to achieve TBC mixing, summarised as follows:

(i)  The first approach is based on Cabibbo-like charged lepton corrections 
$\theta_{12}^e \approx \theta_C$, starting from a zero neutrino mixing angle $\theta_{13}^\nu \approx  0$.
The desired empirical factor of $\sqrt{2}$ in Eqs.\ref{rem1}, \ref{rem2}
then arises automatically from Eq.\ref{1312}, assuming maximal atmospheric neutrino mixing.
The suitable mixing patterns are therefore those with 
$\theta_{13}^{\nu}=0$ and $\theta_{23}^{\nu}=45^o$.
We have considered three such mixing patterns, namely tri-bimaximal (TB) neutrino mixing,
bi-maximal (BM) neutrino mixing, and the Golden Ratio (GR) neutrino mixing,
which each lead to the Sum Rule in Eq.\ref{sumrule}
where $\theta_{12}^{\nu}=35.26^o, 45^o, 31.7^o$, respectively.
Given the prediction $\theta_{13}\approx 9.2^o$, the Sum Rule then yields
a favoured range of $\cos \delta $ in each case, namely $\delta \approx \pm 90^o,
\pm 180^o, \pm 75^o$, respectively. These predictions are testable in future neutrino 
accelerator experiments \cite{Antusch:2007rk}.
We have indicated how such scenarios may be realised in Family Symmetry Models with Pati-Salam symmetry.

(ii) The second approach generates a neutrino mixing angle directly (with no charged lepton corrections), 
$\theta_{13}^\nu  \approx \theta_C/\sqrt{2}$, using 
the type I see-saw mechanism with sequential dominance (SD), assuming 
a particular form of the Dirac neutrino Yukawa couplings in Eqs.\ref{cond1} and \ref{cond2}.
The desired empirical factor of $\sqrt{2}$ in this case 
arises automatically from Eq.\ref{predictions1}, assuming maximal atmospheric neutrino mixing,
and $\Delta \theta_{13}=0$ in Eq.\ref{Delta}, which is satisfied if solar mixing is trimaximal
as follows from Eq.\ref{cond2}. 
The conditions Eqs.\ref{cond1} and \ref{cond2} 
may be justified using family symmetry breaking flavons with particular vacuum alignments in the neutrino sector.
The appearance of $\theta_C$ in the flavon $\varphi_3$ misalignment is justified by the fact that 
$\varphi_3$ is responsible for Cabibbo mixing in the quark sector.
The main prediction of the second approach is that, unlike the first approach, TBC mixing in 
Eqs.~\ref{TBC0}-\ref{TBangles}
is realised accurately, up to corrections of order $\lambda^2$ multiplied by small
coefficients. However, as usual, there will be additional
renormalisation group and canonical normalisation effects which will give additional corrections.

In conclusion, we have proposed the TBC mixing pattern in 
Eqs.\ref{TBC0} and \ref{TBC} and shown how it can be realised
in two very different approaches to quark and lepton mixing, with 
distinctive experimental predictions.

\section*{Acknowledgements}
SFK acknowledges partial support 
from the STFC Consolidated ST/J000396/1 and EU ITN grants UNILHC 237920 and INVISIBLES 289442 .

\end{document}